\documentclass[twocolumn,breaklinks=true]{aastex631}

\usepackage{natbib}
\usepackage{multirow}
\usepackage{gensymb} 
\usepackage[flushleft]{threeparttable} 
\usepackage{amsmath}
\usepackage{amssymb}
\usepackage{url}
\usepackage{wasysym}

\setcitestyle{comma}

\shorttitle{H$\alpha$ at $z\sim8$ from IRAC $5.8\mu$m observations}
\shortauthors{Stefanon et al.}

\begin{document}

\title{High Equivalent Width of H$\alpha$+[N\,II] Emission in $z\sim8$  Lyman-break Galaxies from IRAC $5.8\mu$m Observations: Evidence for Efficient Lyman-continuum Photon production in the Epoch of Re-ionization }

\author{Mauro Stefanon}
\affiliation{Leiden Observatory, Leiden University, NL-2300 RA Leiden, Netherlands}
\affiliation{Departament d'Astronomia i Astrof\'isica, Universitat de Val\`encia, C. Dr. Moliner 50, E-46100 Burjassot, Val\`encia,  Spain}

\author{Rychard J. Bouwens}
\affiliation{Leiden Observatory, Leiden University, NL-2300 RA Leiden, Netherlands}

\author{Garth D. Illingworth}
\affiliation{UCO/Lick Observatory, University of California, Santa Cruz, 1156 High St, Santa Cruz, CA 95064, USA}

\author{Ivo Labb\'e}
\affiliation{Centre for Astrophysics and SuperComputing, Swinburne, University of Technology, Hawthorn, Victoria, 3122, Australia}

\author{Pascal A. Oesch}
\affiliation{Departement d'Astronomie, Universit\'e de Gen\'eve, 51 Ch. des Maillettes, CH-1290 Versoix, Switzerland}
\affiliation{Cosmic Dawn Center (DAWN), Niels Bohr Institute, University of Copenhagen, Jagtvej 128, K\o benhavn N, DK-2200, Denmark}

\author{Valentino Gonzalez}
\affiliation{Departamento de Astronom\'ia, Universidad de Chile, Casilla 36-D, Santiago 7591245, Chile}
\affiliation{Centro de Astrof\'isica y Tecnologias Afines (CATA), Camino del Observatorio 1515, Las Condes, Santiago 7591245, Chile}

\email{Email: stefanon@strw.leidenuniv.nl}

\begin{abstract}

We measure, for the first time, the median equivalent width (EW) of H$\alpha$+[\ion{N}{2}]  in star-forming galaxies at $z\sim8$. Our estimate leverages the unique photometric depth of the \textit{Spitzer}/IRAC $5.8\mu$m-band mosaics (probing $\approx 5500 - 7100$\AA\ at $z\sim8$) of the GOODS Reionization Era Wide Area Treasury from Spitzer (GREATS) program. We median stacked the stamps of $102$ Lyman-break galaxies in the $3.6, 4.5, 5.8$ and $8.0\mu$m bands, after carefully removing potential contamination from neighbouring sources. We infer an extreme rest-frame EW$_0$(H$\alpha$+[\ion{N}{2}])$=2328^{+1326}_{-1127}$\,\AA\ from the measured red $[3.6]-[5.8]=0.82\pm0.27$\,mag, consistent with young ($\lesssim10^7$\,yr) average stellar population ages at $z\sim8$. This implies an ionizing photon production efficiency  of $\log \xi_{\mathrm{ion},0}/\mathrm{erg\ Hz}^{-1}=25.97^{+0.18}_{-0.28}$. Such a high value for photo production, similar to the highest values found at $z\lesssim4$, indicates that only modest escape fractions $f_\mathrm{esc}\lesssim0.3$ (at $2\sigma$) are sufficient for galaxies brighter than $M_\mathrm{UV}<-18$\,mag to re-ionize the neutral Hydrogen at $z\sim8$. This requirement is relaxed even more to $f_\mathrm{esc}\le 0.1$ when considering galaxies brighter than $M_\mathrm{UV}\approx -13$\,mag, consistent with recent luminosity functions and as typically assumed in studies addressing re-ionization. These exceptional results clearly indicate that galaxies can be the dominant source of reionizing photons, and provide us with an exciting glimpse into what we might soon learn about the early universe, and particularly about the Reionization Epoch, from forthcoming \textit{JWST}/MIRI and NIRCam programs. 
\end{abstract}


\section{Introduction}

The characterization of emission lines is a fundamental tool to study the physical processes governing the formation and evolution of galaxies. H$\alpha$ constitutes one of the most reliable estimators of galaxy's star-formation rates (SFRs - e.g., \citealt{moustakas2006, madau2014}) over short timescales  ($\lesssim 10$\,Myr - e.g., \citealt{kennicutt1998, kennicutt2012}) because it tightly correlatates with the production of ionizing photons by OB stars, it does not depend on the metallicity, and it is less affected by dust attenuation compared to rest-UV lines. Moreover, because the rest-frame optical light correlates with the stellar mass ($M_\star$ - e.g., \citealt{stefanon2017b}), the equivalent width (EW) of H$\alpha$ provides a first estimate (modulo a $M_\star/L_\mathrm{optical}$ factor) of the specific star-formation rate (sSFR - e.g., \citealt{fumagalli2012, marmol-queralto2016, faisst2016}).

Optical and NIR spectroscopy have allowed astronomers to probe H$\alpha$ up to $z\sim2.5$ (e.g., \citealt{fumagalli2012, sobral2016, reddy2018, nanayakkara2020}). Progress at $z\sim4-5$ has been enabled by analyzing broad-band photometric data from \textit{Spitzer}/IRAC and interpreting the observed blue $[3.6]-[4.5]<0$\,mag colors as the result of H$\alpha$ emission contributing to the flux density in the $3.6\mu$m band (e.g., \citealt{smit2016, bouwens2016, rasappu2016, faisst2016, caputi2017, lam2019, faisst2019, harikane2018,maseda2020}). 

Constraining H$\alpha$ at $z\gtrsim6$ has proven to be quite challenging for a number or reasons. At these redshifts, the flux densities in both the IRAC $3.6$ and $4.5\mu$m bands are enhanced by nebular line emission ([\ion{O}{3}]+H$\beta$, and H$\alpha$+[\ion{N}{2}], respectively - e.g., \citealt{stefanon2021b}), making it difficult to ascertain whether the observed colors are due to the combination of nebular line and continuum emission, or just to the continuum. This situation is exacerbated by  the general lack of spectroscopic redshifts, essential for identifying which specific nebular lines could be contributing to the flux density in each band. Finally, further uncertainties are introduced by the still unconstrained line ratios at these early epochs  (see e.g., \citealt{brinchmann2008, steidel2014, kewley2015, faisst2016, harikane2018b, stefanon2022a} for discussions on line ratios potentially evolving with cosmic time), and by the significantly shallower ($3-6\times$ depth) data currently available at $5-10\mu$m (e.g., \citealt{stefanon2021c}). These challenges have largely prevented us from securing an emission-line free continuum estimate at rest-frame optical wavelengths.

Fortunately, a favourable window exists again for $7.0\lesssim z \lesssim 8.7$. In this redshift range, H$\alpha$ crosses into the IRAC $5.8\mu$m band, [\ion{O}{3}]+H$\beta$ contribute exclusively to the $4.5\mu$m-band flux density, and the $3.6\mu$m band is free from significant line emission (e.g., \citealt{stefanon2022a}). Thus the potential exists to isolate the key lines to individual bands, and particularly separate H$\alpha$ from significant contamination by other lines.

Notably, this redshift range covers $\approx80\%$ of the Epoch of Reionization (EoR) (see, e.g., \citealt{planck2020_cosmo, robertson2021}). Estimating H$\alpha$ at these epochs, therefore, is particularly valuable to constrain the Lyman-continuum (LyC) ionizing emissivity (e.g., \citealt{leitherer1995}) and the rate of production of H-ionizing photons ($\xi_\mathrm{ion}$, e.g., \citealt{bouwens2016c}).

However, the use of the $5.8$ and $8.0\mu$m bands observations is not a panacea. Both bands suffer from lower sensitivities since $5.8$ and $8.0\mu$m band images could only be acquired during the Spitzer cryogenic mission, whereas those in the $3.6$ and $4.5\mu$m bands continued also during the warm mission. As a result, the sensitivities available in the  $5.8$ and $8.0\mu$m bands are generally $>5-10 \times$ shallower than those available in the $3.6$ and $4.5\mu$m bands. For this reason, most of the studies characterizing the physical properties of EoR galaxies have so far focused on estimating the intensity of  [\ion{O}{3}]+H$\beta$ from the measured $[3.6]-[4.5]$ color (\citealt{smit2014, castellano2017, debarros2019, stefanon2019,stefanon2022a, bowler2020, strait2020, strait2021,endsley2021}). Indeed, no detections of H$\alpha$ from the color excess in the $5.8\mu$m band have been published so far.  The few estimates existing in the literature are indirect, and stem from converting the EW([\ion{O}{3}]+H$\beta$) assuming standard Case B recombination coefficients and metal-line ratios either from best-fit SED analyses or extracted from tabulated values (e.g., \citealt{smit2014, stefanon2022a, endsley2021}). Recently, attempts to detect emission in the $5.8\mu$m band for individual $z\sim7-8$ Lyman-break galaxies (LBGs) were performed by \citet{asada2022}, leveraging the lensing magnification of the foreground galaxy clusters  Abell2744, Abell1063, Abell370, and MACS-J0717 from the Hubble Frontier Fields program (HFF - \citealt{lotz2017}) were not successful. The lack of direct detections has resulted in a chronic absence of direct constraints on the H$\alpha$ intensity at these pivotal redshifts.

One possible approach to compensate for the current lack of deep $5-10\mu$m data consists in combining the imaging available for samples of galaxies, and extracting their average properties. Our recently released IRAC mosaics from the GOODS Re-ionization Era wide-Area Treasury from Spitzer program (GREATS, PI: I. Labb\'e - \citealt{stefanon2021c}) include all the relevant IRAC observations acquired in the four bands over the CANDELS (\citealt{grogin2011, koekemoer2011}) GOODS-N and GOODS-S fields (\citealt{giavalisco2004}) across the almost $2$ decades of Spitzer operations. Notably for this study, the GREATS $5.8\mu$m-band mosaic provides $\approx4-140 \times$ deeper coverage than what exists for the  Abell2744, Abell1063, Abell370, and MACS-J0717 HFF fields\footnote{\url{https://irsa.ipac.caltech.edu/data/SPITZER/Frontier/overview.html}}  and $\gtrsim1.5 \times$ deeper coverage over a $\gtrsim4\times $ larger area than the IRAC Dark Field\footnote{\url{http://web.ipac.caltech.edu/staff/jason/darkfield/index.html}} (\citealt{krick2009}).  The GREATS mosaics therefore constitute the deepest and thus most suitable data set for probing H$\alpha$ emission in $z\sim8$ galaxies prior to JWST operations. 

In this study, we explore for the first time the intensity of the H$\alpha$ emission in $z\sim8$ galaxies by stacking the image stamps in the IRAC bands centered on $102$ candidate LBGs at $7.3<z_\mathrm{phot}<8.7$ identified by \citet{bouwens2015} in the CANDELS GOODS, UDS and COSMOS fields. This sample was already utilized by \citet{stefanon2022a} to study the rest-frame optical properties of $z\sim8$ galaxies, and benefits from minimal neighbour contamination (see Section \ref{sect:sample} and \citealt{stefanon2022a}).

The layout of this paper follows: in Section \ref{sect:sample} we present the data set and the sample adopted in this study; in Section \ref{sect:stacking} we describe the procedure we followed to estimate the average flux densities; the stacked photometry and the estimate of the rest-frame EW$_0$(H$\alpha$) are presented in Section \ref{sect:results}; we place our results in the context of the evolution of the EW$_0$(H$\alpha$), sSFR and $\xi_\mathrm{ion}$ in Section \ref{sect:discuss}. A summary of this study together with our conclusions is presented in Section \ref{sect:conclusions}.

Throughout this paper, we adopt $\Omega_M=0.3$, $\Omega_\Lambda=0.7$ and $H_0=70$\,km s$^{-1}$ Mpc$^{-1}$, consistent with the most recent estimates from Planck (\citealt{planck2020_cosmo}). Magnitudes are given in the AB system (\citealt{oke1983}), while our $M_\star$ and SFR measurements are expressed in terms of the \citet{salpeter1955} initial mass function (IMF). For brevity, we denote the \textit{HST} F435W, F606W, F775W, F850LP, F105W, F125W, F140W and F160W as $B_{435}$, $V_{606}$ and $i_{775}$, $z_{850}$, $Y_{105}$, $J_{125}$, $JH_{140}$ and $H_{160}$.

\begin{figure*}
\includegraphics[width=18cm]{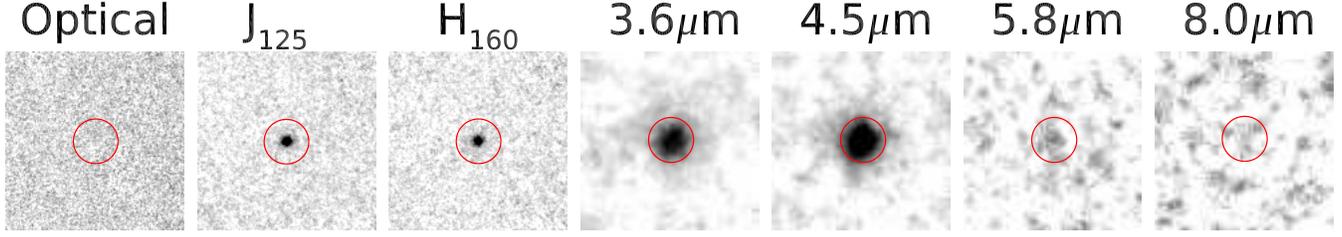}
\caption{Image stamps ($\sim8\farcs0$ per side) in the IRAC and \textit{HST} bands, centered on the median stacks. The red circle marks the $2\farcs0$ diameter aperture adopted for the photometry of the IRAC stacks.  The \textit{HST} stacks are presented to provide a better visual context of the data involved in our study, as the median flux densities in \textit{HST} bands were estimated from the photometry of individual sources. Each stamp refers to a different band, as labeled at the top; in particular the \textit{HST} optical stack combines all data available in the $B_{435}, V_{606}, i_{775}$ and $z_{850}$ bands. Remarkably, we find a $4.3\sigma$ detection for the flux density in the $5.8\mu$m band, while a $\sim1.8\sigma$ measurement in the similarly-deep $8.0\mu$m stack. The striking visual difference between the detection in the $5.8\mu$m band compared to those in the $3.6$ and $4.5\mu$m bands is a direct consequence of the $\sim 40\times$ lower sensitivity available in the $5.8\mu$m band. Our measurement suggests a significant contribution from $H\alpha$ emission to the flux density in the $5.8\mu$m band, as discussed in Section \ref{sect:results}. \label{fig:stack_stamps}}
\end{figure*}

\begin{figure}
\includegraphics[width=9cm]{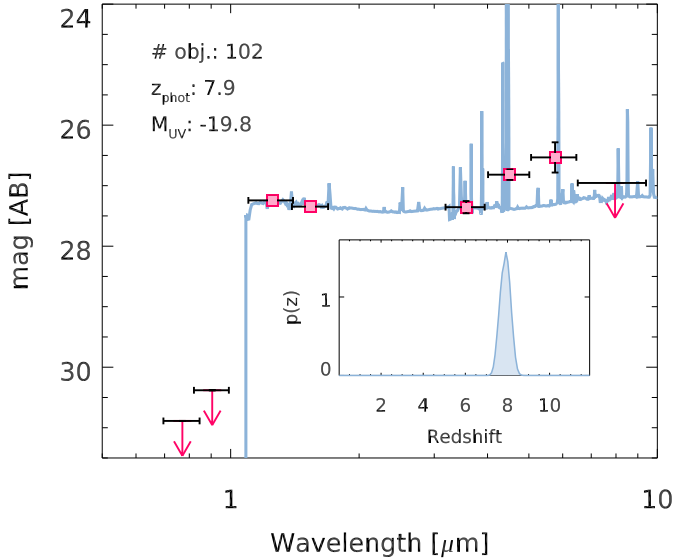}
\caption{Median-stacked SED resulting from our analysis. The filled red squares with error bars mark the stacked photometry, while the red arrows represent $2 \sigma$ upper limits. The black horizontal bars indicate the effective width of the bands.  The blue curve corresponds to the best-fitting \textsc{EAzY} template. The inset presents the redshift probability distribution computed by \textsc{EAzY}. The labels at the top-left corner present the number of objects entering the stack, the median redshift, and the $M_\mathrm{UV}$ computed by $\textsc{EAzY}$. Remarkably, the flux density in the $5.8\mu$m band is comparable to that in the $4.5\mu$m band, $\sim2.2\times$ higher than that in the \textit{HST} and IRAC $3.6\mu$m bands, indicative of strong nebular line emission from $H\alpha$. Also evident again is the lack of a prominent Balmer break between the $H_{160}$ and $3.6\mu$m bands, a result previously emphasised in \citet{stefanon2022a}, the lead-up study to this analysis.  \label{fig:SED}}
\end{figure}

\begin{deluxetable*}{llccccccc}
\tablecaption{Observational data used for the SMF estimates. \label{tab:obs_data}}
\tablehead{\multicolumn{2}{c}{Field} & \colhead{Area} & \colhead{$H_\mathrm{160}$\tablenotemark{a}} &\colhead{IRAC Data\tablenotemark{b}} & \colhead{$3.6\mu$m\tablenotemark{c}} & \colhead{$4.5\mu$m\tablenotemark{c}} & \colhead{$5.8\mu$m\tablenotemark{c}} & \colhead{$8.0\mu$m\tablenotemark{c}}    \\
\multicolumn{2}{c}{Name} & \colhead{[arcmin$^2$]} & \colhead{$5\sigma$ [mag]} & & \colhead{$5\sigma$ [mag]} & \colhead{$5\sigma$ [mag]}  & \colhead{$5\sigma$ [mag]} & \colhead{$5\sigma$ [mag]}  
}
\startdata
\multicolumn{2}{l}{XDF}      & $4.7$   & $29.4$ & GREATS        & $\sim27.2$  & $\sim26.7$   & $\sim23.9$ & $23.5-23.8$ \\
\multicolumn{2}{l}{HUDF09-1} & $4.7$   & $28.3$ & GREATS        & $\sim26.3$  & $\sim25.8$   & $\sim23.7$ & $\sim22.2$  \\
\multicolumn{2}{l}{HUDF09-2} & $4.7$   & $28.7$ & GREATS        & $\sim27.0$  & $25.5-26.0$  & $\sim22.5$  & $\sim22.2$ \\
\multicolumn{2}{l}{ERS}      & $40.5$  & $27.4$ & GREATS        & $26.2-27.0$ & $25.6-26.7$  & $\sim23.5$ & $\sim23.3$ \\
CANDELS & GOODS-N Deep       & $62.9$  & $27.5$ & GREATS        & $27.0-27.3$ & $26.5-26.8$ & $23.5-24.3$ & $23.3-24.0$  \\
        & GOODS-N Wide       & $60.9$  & $26.7$ & GREATS        & $26.3-27.2$ & $25.8-26.8$ & $23.5-24.1$  & $23.3-23.9$\\
        & GOODS-S Deep       & $64.5$  & $27.5$ & GREATS        & $\sim27.3$  & $26.6-26.9$ & $23.5-23.8$ & $23.3-23.8$\\
        & GOODS-S Wide       & $34.2$  & $26.8$ & GREATS        & $26.5-27.2$ & $26.2-26.7$  & $\sim23.5$  & $\sim23.3$\\
        & COSMOS             & $151.9$ & $26.8$ & SEDS+SCANDELS & $26.4-26.7$ & $26.0-26.3$  & $\sim21.2$  & $\sim21.0$ \\
        &              &  &  & +S-COSMOS &  &   &   &  \\
        & UDS                & $151.2$ & $26.8$ & SEDS+SCANDELS & $25.4-26.3$ & $25.0-25.9$  & $\sim21.5$  & $\sim21.7$ \\
        &              &  &  & +SpUDS &  &   &   &  \\[5pt]
\hline
\multicolumn{2}{c}{Totals:} & $580.2$ & & & &  \\
\enddata
\tablenotetext{a}{$5\sigma$ limit from \citet{bouwens2015}, computed from the median of measured uncertainties of sources.}
\tablenotetext{b}{GREATS: \citet{stefanon2021c}; SEDS: \citet{ashby2013a}; SCANDELS: \citet{ashby2015}; S-COSMOS:  \citet{sanders2007}; SpUDS: \citet{caputi2011}.}
\tablenotetext{c}{Nominal $5\sigma$ limit for point sources from the SENS-PET exposure time calculator, based on the exposure time maps. Due to inhomogeneities in the coverage, a range of values is quoted when the depth varies by more than $\sim0.2$\,mag across the field. Because of the combined effects of broad \textit{Spitzer}/IRAC PSF and significant exposure times, source blending may reduce the actual depth (see discussion in \citealt{labbe2015} and \citealt{stefanon2021c}).}
\end{deluxetable*}

\section{Data and Sample}
\label{sect:sample}

The sample adopted for this study consists of the $102$  candidate $z\sim8$ Lyman-break galaxies  previously discussed in \citet{stefanon2022a}. Briefly, this sample is based on the $Y-$dropout LBGs that \citet{bouwens2015} identified over the CANDELS (\citealt{grogin2011, koekemoer2011}) GOODS-N, GOODS-S (\citealt{giavalisco2004}), UDS (\citealt{lawrence2007}) and COSMOS (\citealt{scoville2007}) fields, the ERS field (\citealt{windhorst2011}), and the UDF/XDF (\citealt{beckwith2006,illingworth2013, ellis2013}) with the HUDF09-1 and HUFD09-2 parallels (\citealt{bouwens2011b})\footnote{We excluded CANDELS/EGS because of the lack of deep data in the Y band, which makes the selection of candidate $z\sim8$ LBGs more uncertain.}.  In Table \ref{tab:obs_data} we summarize the main properties of the adopted data sets. The mosaics are characterized by $5 \sigma$ depths of $\approx27.5$~mag in the $V_{606}$ and $I_{814}$ bands,  $\approx26.7-27.5$~mag in the $Y_{105}$ (GOODS fields) and $26.0$~mag in the ground-based $Y$ band (UDS and COSMOS), and $\sim26.8-27.8$\,mag in the $J_{125}$ and $H_{160}$ bands. 

A crucial aspect for this study is that these fields have excellent coverage in the \textit{Spitzer}/IRAC $3.6, 4.5, 5.8$ and $8.0\mu$m bands. In particular, for the GOODS fields we adopted the mosaics and location-dependent point-spread functions (PSFs) from the GREATS program (PI: Labb\'e - \citealt{stefanon2021c}). These mosaics combine all the useful IRAC data acquired across the full scientific life of \textit{Spitzer}. As a result they are very deep, with $5 \sigma$ depths of  $\sim26.0-27.0$\,mag in the IRAC $3.6$ and $4.5\mu$m bands, and  $\sim 23.0-24.0$\,mag in the IRAC $5.8$ and $8.0\mu$m bands. While deep, we also require accurate PSFs to minimize the contamination from neighbours (see e.g., \citealt{stefanon2021c}). Given the asymmetric nature of the instrumental IRAC PSF, particularly in the $3.6$ and $4.5\mu$m bands, and the variety of programs included in the mosaics, the PSFs can significantly vary across each field. The PSFs for GREATS are reconstructed by combining a high S/N empirical template rotated according to the position angle and weighted through the coverage depth from each program at the specific location. 

We validated the redshift of the individual sources in our sample by running \textsc{EAzY} \citep{brammer2008} on the set of \textit{HST} observations available for each source, while requiring $7.3\le z_\mathrm{phot} \le 8.7$. At these redshifts, the IRAC colors are sensitive to the intensity of the main rest-frame optical emission lines. A number of studies have shown that the IRAC colors can be successfully used to significantly reduce the photometric redshift uncertainties (e.g., \citealt{smit2014, roberts-borsani2016}). However, this could also potentially bias our sample towards sources with strong emission lines. For this reason, we excluded the flux densities in the IRAC bands when running \textsc{EAzY}. Reassuringly, inclusion of IRAC fluxes in estimating the photometric redshifts has no strong impact on the sources we select ($91$ sources, corresponding to $\sim89\%$, are in common between the two samples).

Because of the broad point-spread function of IRAC ($\approx1\farcs5-2\farcs0$ from the $3.6\mu$m to the $8.0\mu$m band - \citealt{stefanon2021c}), the extended light profiles of neighbouring objects could systematically affect the measurement of the emission of specific sources. For this reason, in our analysis we subtracted the neighbour emission with \textsc{Mophongo} \citep{labbe2006, labbe2010a, labbe2010b, labbe2013, labbe2015}, and removed from the sample those sources where visual inspection still showed residual contamination  \citep{stefanon2022a}.\\

To allow for a more meaningful comparison of our results with the literature, we also estimated the main stellar population parameters. These were computed by running  \textsc{FAST} \citep{kriek2009} on the \citet{bruzual2003} spectral energy distribution (SED) templates for $Z_\mathrm{star}=0.2Z_\odot$ metallicity with \citet{salpeter1955} IMF, a constant star-formation history, and a \citet{calzetti2000} extinction law. The set of templates was post-processed through \textsc{Cloudy} version 17.02 (\citealt{ferland2017}), assuming a spherical constant-density nebula with $n(H)  = 100$\, cm$^{-3}$, $Z_\mathrm{gas}=0.2Z_\odot$ metallicity, an ionization parameter $\log U = -2.5$  (e.g., \citealt{stark2017, debarros2019}), and a negligible escape fraction.

\section{Stacking}
\label{sect:stacking}

Following \citet{stefanon2022a}, we adopted distinct stacking procedures for the \textit{HST} and for the IRAC bands, given the different photometric depths. For the \textit{HST} bands, stacking consisted in evaluating the median of the extracted photometry normalized by the flux density in the $H_{160}$ band of each source, as the generally higher S/N characterizing these data reduces the measurement scatter around the true value. For the IRAC bands, however, the lower S/N compared to the  \textit{HST} data could introduce a larger scatter in the final measurement, possibly even systematically affecting it. We therefore constructed image stacks by taking the median of the image stamps centered on each source after they have been cleaned from neighbours using \textsc{Mophongo} and normalized by the $H_{160}$ flux density of each source. The stacked IRAC flux densities were measured in $2\farcs0$-diameter apertures. The smaller aperture adopted here compared to what \citet{stefanon2022a} used is a trade-off between optimizing the S/N and minimizing potential flux loss introduced by the challenges in aligning the sources before taking the median and removal of neighbour contamination, particularly in the $5.8$ and $8.0\mu$m data. The aperture photometry was corrected to total using the median of the PSFs reconstructed at the location specific to each source. The applied correction factors are $\sim2.2, 2.2, 2.9$ and $3.3$ for the $3.6, 4.5, 5.8$ and $8.0\mu$m bands, respectively. Uncertainties associated with the flux densities were computed  by bootstrapping the sample $1000$ times. Finally, all values were rescaled by the median of the flux densities in the $H_{160}$ band. An analysis adopting larger apertures ($2\farcs5$ and $3\farcs6$) resulted in measurements consistent at $1\sigma$ with those obtained with the smaller aperture, albeit with larger uncertainties.

We further validated our $5.8\mu$m-band measurement through a Monte Carlo simulation, presented in Appendix \ref{app:simul}. Briefly, we applied the same neighbour removal and stacking procedure we adopted for our main analysis to $102$ synthetic sources. We added them to the IRAC $5.8\mu$m mosaics, after normalizing their flux densities to those expected for the LBG in our sample, assuming a flat $f_\nu$ SED and a rest-frame EW$_0$(H$\alpha$)$=1900$\AA. This whole process was repeated $100$ times. The resulting distribution of flux density measurements shows that, on average, we can recover the input flux density and that the impact of possible contamination by neighbours is negligible, as discussed and shown in more detail in the Appendix and in Figure \ref{fig:simul}. \\

The depth of the IRAC $5.8$ and $8.0\mu$m mosaics in the COSMOS and UDS fields is $\sim2$\,mag shallower (corresponding to $\sim6\times$ brighter flux limits) than the average depth in the GOODS fields, suggesting we should perhaps exclude or de-weight them in our stacks. On the other hand, these two CANDELS-Wide fields do allow us to incorporate $10$ of the brightest $z\sim8$ sources with deep \textit{HST} imaging from the \citet{bouwens2015} catalogs, providing a more comprehensive view of the properties of $z\sim8$ galaxies. Even so, the median brightness of the sources selected in the COSMOS and UDS fields is only a factor $\sim3$ brighter than the median for the sources in the GOODS fields. To evaluate the impact of these sources on our stack results, we repeated our stacking analysis excluding the $10$ sources in the COSMOS and UDS fields. Reassuringly, the flux densities of the new measurements in the $3.6, 4.5$ and $5.8\mu$m bands differ by $\sim 5-10\%$ from those obtained with full sample, after the change in $H_{160}$ normalization is taken into account. The flux density in the $8.0\mu$m band is $\sim60\%$ fainter but still consistent at the $1\sigma$ level with that from the full sample.  On balance, we therefore opted for including in our stack the $z\sim8$ galaxies from the UDS and COSMOS fields.

\section{Results}
\label{sect:results}

\begin{deluxetable*}{lccccccccc}
\tablecaption{Flux densities for our median-stacked photometry \label{tab:stack_phot}}
\tablehead{ & \colhead{$V_{606}$} & \colhead{$i_{775}$} & \colhead{$z_{850}$}  & \colhead{$J_{125}$} & \colhead{$H_{160}$}  & \colhead{$3.6\mu$m} & \colhead{$4.5\mu$m} & \colhead{$5.8\mu$m} & \colhead{$8.0\mu$m}       \\
 & \colhead{(nJy)}  & \colhead{(nJy)} & \colhead{(nJy)} & \colhead{(nJy)} & \colhead{(nJy)} & \colhead{(nJy)} & \colhead{(nJy)} & \colhead{(nJy)} & \colhead{(nJy)}  
}
\startdata
Stack & $-0.4 \pm  0.7$ & $-0.5 \pm  0.8$ & $-1.3 \pm  1.3$ & $46.0 \pm  2.4$ & $41.8 \pm  1.7$ & $41.5 \pm  3.9$ & $68.0 \pm  5.4$ & $88.5 \pm 20.3$ & $56.3 \pm 30.0$ \\
\enddata
\tablecomments{We only list the flux densities in those bands available for at least $90\%$ of the sources in our sample.}
\end{deluxetable*}

The stacked stamps in the IRAC bands are presented in Figure \ref{fig:stack_stamps}, while the photometry in those bands offering coverage for at least $90\%$ of the sources in our sample is listed in Table \ref{tab:stack_phot}, and displayed in Figure \ref{fig:SED}. Our photometric measurements are characterized by  $\gtrsim20\sigma$ detections in the \textit{HST} $J_{125}$ and $H_{160}$ bands, and  $\sim10\sigma$ in the $3.6$ and $4.5\mu$m bands. Remarkably, the stack in the IRAC $5.8\mu$m band has resulted in a $\sim4.3\sigma$ detection, while the stack in the $8.0\mu$m band is characterized by a $1.8\sigma$ significance. Because the $5.8\mu$m- and $8.0\mu$m-band mosaics adopted in our study have similar depths and image quality, the $8.0\mu$m band detection, even though of somewhat lower S/N, actually provides valuable added support for our $4.3\sigma$ measurement at $5.8\mu$m as being a genuine detection.  Together these detections give added confidence that the detection is real and is not significantly affected by neighbour and/or interloper contamination.

To assist in the interpretation of our measurements and to further validate the consistency of the stacked photometry, we also present the best-fit SED template from \textsc{EAzY} in Figure \ref{fig:SED}. For this step, we complemented the default set with SEDs of young (age$\sim10^{6-8}$\,yr) star-forming galaxies from BPASS v1.1 \citep{eldridge2017}, whose nebular emission was computed with \textsc{Cloudy} (\citealt{ferland2013, ferland2017}). A formal fit with \textsc{FAST} \citep{kriek2009} adopting our default configuration (see Section \ref{sect:sample}) results in a stellar mass of $M_\star=10^{8.12^{+0.86}_{-0.28}}M_\odot$, a dust extinction of $A_V=0.2^{+0.1}_{-0.2}$\,mag and a stellar population age of $\log(\mathrm{age/yr})=7.1^{+1.0}_{-0.5}$. These properties are consistent with the average of the properties \citet{stefanon2022a} found for $z\sim8$ LBGs from several samples binned by UV luminosity. \\

The most notable feature evidenced by our stacked photometry is the red $[3.6]-[5.8]=0.82\pm 0.27$\,mag color. This is in addition to the lack of a Balmer break between the $H_{160}$ and $3.6\mu$m band and the robustly red $[3.6]-[4.5]$ colors, indicative of strong [\ion{O}{3}]+H$\beta$ emission.  Both these results have already been discussed in detail in \citet{stefanon2022a}. The $[3.6]-[5.8]$ color was more poorly constrained in the study of \citet{stefanon2022a} due to the lower S/N in the $5.8\mu$m band likely resulting from splitting the sample across four luminosity bins. At $z\sim6.5-8.9$, a number of optical emission lines fall within the $5.8\mu$m band, with H$\alpha$ and [\ion{N}{2}] being expected to contribute the most to the flux measurements (e.g.\citealt{anders2003}). Under the assumption that the flux density in the $5.8\mu$m band arises from the combination of H$\alpha$ and [\ion{N}{2}] with the stellar and nebular continuum, the measured $[3.6]-[5.8]$ color corresponds to a rest-frame EW$_0$(H$\alpha$+[\ion{N}{2}])$=2328^{+1326}_{-1127}$\AA.  For the EW$_0$(H$\alpha$) alone we derive an extreme EW$_0$(H$\alpha$)$=1960^{+1089}_{-927}$\,\AA, after iteratively accounting for the contribution of [\ion{O}{2}] and other less prominent lines in the $3.6\mu$m band, assuming the line ratios of \citet{anders2003} for a $Z=0.2Z_\odot$ metallicity and negligible dust extinction. This estimate corresponds to a luminosity $\log(L_{\mathrm{H}\alpha}/[\mathrm{erg\, s}^{-1}])=42.62^{+0.15}_{-0.23}$ and a SFR=$36_{-15}^{+14}M_\odot$yr$^{-1}$ \citep{kennicutt2012}. 

To our knowledge, this measurement constitutes the first detection of H$\alpha$ from broad-band photometry in \textit{normal} star-forming galaxies at $z>6.5$. Recent attempts at measuring the intensity of H$\alpha$ at $z\sim8$ for individual sources in HFF cluster fields were unsuccessful (e.g., \citealt{asada2022}), likely due to the shallow coverage available in the IRAC $5.8\mu$m band over those fields combined with the relatively low magnification values for the considered sources. 

Such an elevated EW$_0$(H$\alpha$) could originate from active galactic nuclei (AGN). Indeed, indication of nuclear activity in $z\sim8$ galaxies has been recently found (e.g., \citealt{laporte2017,mainali2018,topping2021}). However, the extrapolation to $z\sim8 $ of recent results at $z\lesssim7$ (e.g., \citealt{harikane2022}) suggests that AGN would be a marginal population in the $L<L^*$ galaxies which dominate our sample. A more definitive assessment of the fraction of AGN in sub-$L^*$ galaxies at $z\sim8$ requires spectroscopic data, still unavailable for statistically significant samples.

The stacked SED also shows a red $[3.6]-[4.5]=0.54\pm0.13$\,mag color. The increased flux density in the $4.5\mu$m band is likely to result from substantial enhancement by [\ion{O}{3}]$_{\lambda\lambda 4959,5007}$ and H$\beta$ line emission. Applying the same measurement procedure adopted for the estimate of the EW$_0$(H$\alpha$), the measured $[3.6]-[4.5]$ color corresponds to an EW$_0$([\ion{O}{3}]+H$\beta$)$=1006^{+230}_{-220}$\,\AA. This value implies an EW$_0$(H$\alpha$)$=697^{+160}_{-153}$\,\AA. Our more direct measurement of  EW$_0$(H$\alpha$) based on the $5.8\mu$m band excess differs by only $\sim1.3\sigma$ from this estimate, providing further confirmation that H$\alpha$ is very strong in these $z\sim8$ LBGs. Finally, our stacked SED is characterized by a $J_{125}-H_{160}=-0.10\pm0.07$\,mag color, indicating a blue UV slope ($\beta\sim -2.4$), and a flat $H_{160}-[3.6]=-0.01\pm0.11$\,mag suggesting young stellar population ages. \citet{stefanon2022a} already provide an extensive discussion of the interpretation of stack results involving these bands ($J_{125}, H_{160}, 3.6\mu$m, and $4.5\mu$m).

\begin{figure*}
\centering\includegraphics[width=12cm]{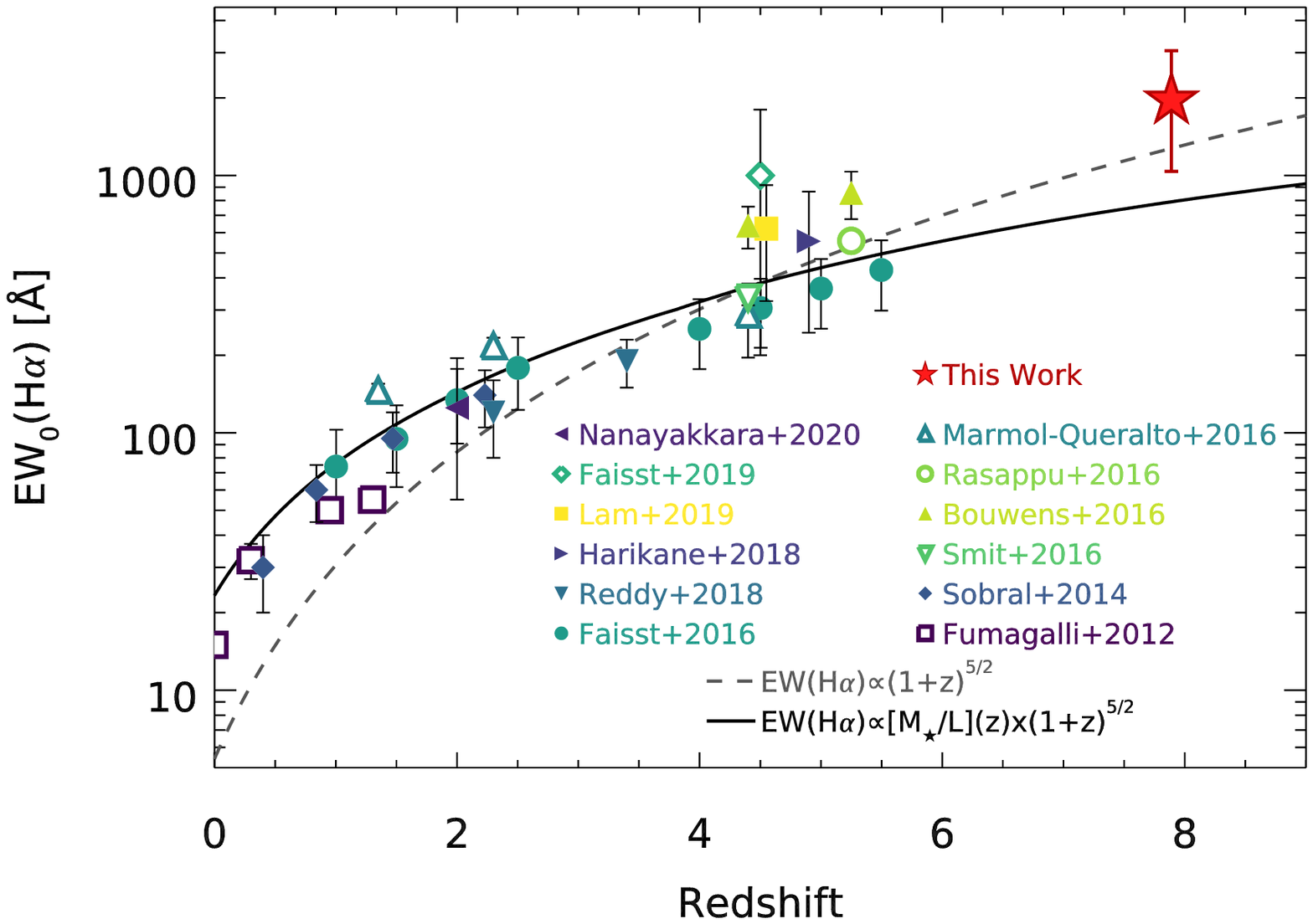}
\caption{Evolution of the rest-frame EW$_0$(H$\alpha$) since $z\sim8$. Our new $z\sim8$ measurement is shown relative to other measurements from the literature for $z<6$, as indicated by the legend.  Specifically, we included estimates from \citet{fumagalli2012, sobral2014,smit2016, bouwens2016, rasappu2016, marmol-queralto2016, faisst2016, reddy2018, harikane2018, lam2019, faisst2019} and \citet{nanayakkara2020}. The dashed grey curve marks the EW$_0$(H$\alpha$) expected when its evolution follows that of the sSFR under a non-evolving star-formation efficiency scenario  (e.g., \citealt{dekel2013}) and constant $M_\star/L_\mathrm{optical}$ ratio, while the solid black curve is for a $M_\star/L_\mathrm{optical}$ ratio that increases with decreasing redshift according to a constant star-formation history.  \label{fig:EWHa}}
\end{figure*}

\section{Discussion}
\label{sect:discuss}

\subsection{Evolution of the EW$_0$(H$\alpha$)}

The large EW$_0$(H$\alpha$) we infer requires very young stellar populations, $\lesssim$few$\times10^{7}$\,yr (e.g., \citealt{inoue2011, wilkins2020}), consistent with our age estimate based on multi-band photometry and with recent measurements at similar epochs (e.g., \citealt{stefanon2022a, endsley2021, strait2020}, but see e.g., \citealt{roberts-borsani2020, roberts-borsani2022,tacchella2022}).

In Figure \ref{fig:EWHa} we compare the EW$_0$(H$\alpha$) estimate from this study to measurements at $z<7$ from the literature. Specifically, we considered the estimates from \citet{fumagalli2012, sobral2014,smit2016, bouwens2016, rasappu2016, marmol-queralto2016, faisst2016, reddy2018, harikane2018, lam2019, faisst2019} and \citet{nanayakkara2020}. We included only measurements with $M_\mathrm{UV}$ within $\sim\pm0.75$\,mag of $M_\mathrm{UV}=-19.9$\,mag, or whose stellar masses $M_\star$ lie within $\pm1$\,dex of the stellar mass we estimated from our stacked photometry, when a typical $M_\mathrm{UV}$ was not quoted with a result. Our estimate constitutes one of the highest EW$_0$(H$\alpha$) measurements across the $0<z<8$ redshift range. Nonetheless, the present large uncertainty of our measurement make it consistent at $\sim1.5\sigma$ with the average EW$_0$(H$\alpha$) existing at $z\sim4-6$.

Given that the samples in Figure \ref{fig:EWHa} were comparably selected over the full redshift range, we can also address the question of the specific star-formation rate (sSFR) over this wide time baseline.  At first approximation the EW(H$\alpha$) is proportional to the sSFR, and so we can also use the trends seen in Figure \ref{fig:EWHa} to characterie the evolution with redshift of the sSFR. This is shown in the form sSFR$\propto(1+z)^{5/2}$, as derived by  \citet{dekel2013}, applying a constant conversion factor to the analytical expression of the evolution of the specific accretion rate of the dark matter halo under the hypothesys of a non-evolving ratio between the stellar mass and the mass of the host dark matter halo. We applied an overall normalization by fitting the curve to the observations. This simple relation can reproduce well the observations at $z\gtrsim4$, but underestimates the expected EW(H$\alpha$) at lower redshifts, with larger gaps for lower redshift values. 

While the uncertainty on the present EW$_0$(H$\alpha$) measurement is large, one possible explanation for the differential evolution observed between the sSFR and the EW$_0$(H$\alpha$) is a $M_\star/L_\mathrm{optical}$ ratio (where $L_\mathrm{optical}$ refers to the luminosity of the continuum at wavelengths close to that of H$\alpha$) evolving with cosmic time. This is indeed expected considering the increasingly larger fractions of evolved stellar populations at later cosmic times. The black solid curve in Figure \ref{fig:EWHa} presents the result of applying a redshift-dependent $M_\star/L_\mathrm{optical}$ ratio to the analytical expression of the sSFR($z$) evolution. This factor was estimated from our default template set (Section \ref{sect:sample}), assuming galaxies started forming stars at $z\sim20$ (e.g., \citealt{mawatari2020,harikane2022})\footnote{The exact burst redshift does not significantly influence our conclusions because the time difference between $z\sim20$ and e.g., $z\sim15$ is just $<100$Myr}. We applied a global normalization factor from fitting the curve to the available observations. Here we do not consider the effects of dust attenuation given the growing indication that at the stellar masses considered here they are not a significant factor at $z>2$ (e.g., \citealt{bouwens2016b,dunlop2017, mclure2018, bouwens2020}) and marginal at $z<2$ (e.g., \citealt{garn2010}). The curve matches the observations reasonably well, in particular for $z\lesssim6$. Our measurement  at $z\sim8$ is consistent  at $\gtrsim1\sigma$ with the values expected from the new relation, although this is due, at least in part, to the large uncertainties. Nonetheless, the overall agreement in the recovered EW$_0$(H$\alpha$) with the sSFR evolution to $z\sim8$ supports a scenario of a marginally evolving star-formation efficiency, as suggested by recent observational studies (e.g., \citealt{stefanon2017b,oesch2018,harikane2018, bouwens2021, stefanon2021b, stefanon2022a}).

\begin{figure*}
\centering\includegraphics[width=12cm]{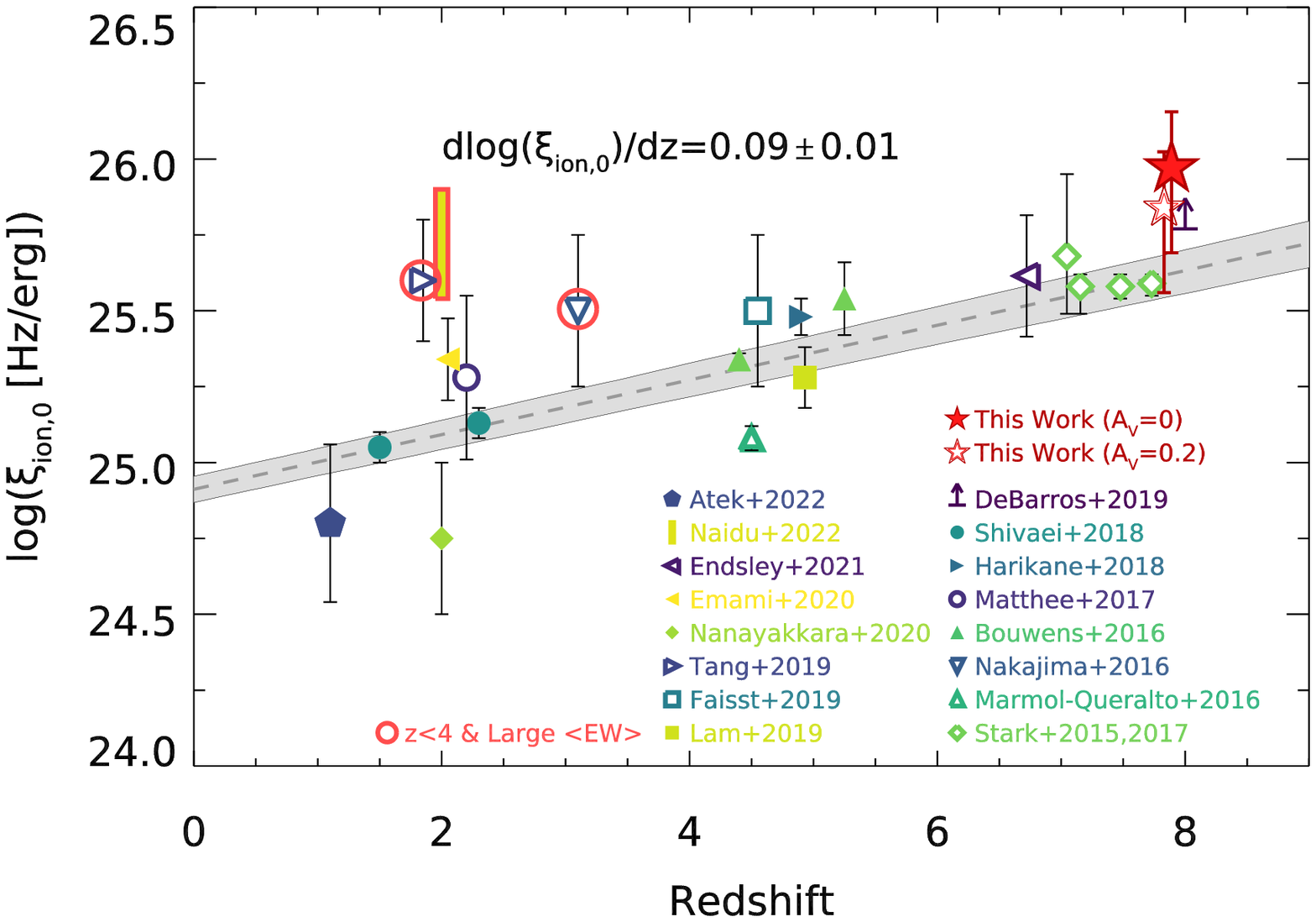}
\caption{Evolution of $\xi_\mathrm{ion}$ for $1.5\lesssim z\lesssim 8 $. The compilation of measurements, listed in the legend, includes the results of \citet{stark2015, stark2017, marmol-queralto2016,nakajima2016, bouwens2016, matthee2017, harikane2018, shivaei2018,debarros2019,lam2019,faisst2019,tang2019,nanayakkara2020,emami2020,endsley2021b, naidu2022} and \citet{atek2022}. We only considered measurements corresponding to $M_\mathrm{UV}\sim-19.9$\,mag or $\log(M_\star/M_\odot)\sim 8.1\pm1.0$ when the $M_\mathrm{UV}$ information was missing. The red open circles flag those results at $z<4$ whose sample was explicitly selected to have rest-optical lines with EW$_0\gtrsim1000$\,\AA\ (\citealt{nakajima2016, tang2019}) or with likely a hard ionizing spectra (\citealt{naidu2022}). We arbitrarily shifted by $\Delta z=-0.06$ our measurement for $A_V=0.2$\,mag (open star) to improve the readability. The dashed line marks the result of a linear fit, with the $68\%$ confidence interval encompassed by the grey shaded area. This composite set of measurements suggests a steady increase of $\xi_\mathrm{ion}$ with increasing redshift.  \label{fig:csi_z}}
\end{figure*}

\subsection{Implications of the high EW$_0$(H$\alpha$) on re-ionization}

\subsubsection{Constraints on $\xi_\mathrm{ion}$}

We can now use our new determination of the EW$_0$(H$\alpha$) to derive the efficiency of production of H-ionizing photons ($\xi_\mathrm{ion}$). This enables quantifying a key parameter, the total ionizing power of galaxies, in the heart of the reionization epoch and close to the time of instantaneous reionization ($z=8.8$, \citealt{planck2016_reion}). This is particularly valuable since only few, less direct, measurements exist at $z>7$ (.e.g., \citealt{stark2015,stark2017,debarros2019, endsley2021b}), inferred from either spectral analysis of rest-UV emission lines or from SED fitting to broad-band photometry.

Following \citet{bouwens2016c}, we compute $\xi_\mathrm{ion}$ from the production rate of Lyman-continuum photons, $\dot N(H^0)$. This can be inferred from the H$\alpha$  luminosity $L({\mathrm{H}\alpha})$, using the relation of \citet{leitherer1995}:

\begin{equation}
L(\mathrm{H}\alpha) = 1.36 \times 10^{-12}\dot N(H^0)
\end{equation}
\noindent where $L(\mathrm{H}\alpha)$ has units of erg s$^{-1}$ and $\dot N(H^0)$ of s$^{-1}$. This relation has a small ($\lesssim 15\%$ or $0.06$\,dex) dependence on the metallicity and electron temperature  (e.g., \citealt{charlot2001}), which we assume as systematic uncertainty.  This uncertainty is  significantly smaller than the stochastic uncertainties associated with the EW$_0$(H$\alpha$) measurement. The LyC photon production efficiency $\xi_{\mathrm{ion},0}$ (where the subscript $0$ indicates the assumption of an escape fraction $f_\mathrm{esc}=0$, i.e., this is the actual production rate, in the galaxy, excluding any losses) can then be computed as:

\begin{equation}
\xi_{\mathrm{ion},0}=\frac{\dot N(H^0)}{L_\mathrm{UV}}
\end{equation}

\noindent where $L_\mathrm{UV}$ is the UV-continuum luminosity computed from the stacked SED. The application of the above relations to our measurements yields $\xi_{\mathrm{ion},0}=10^{25.97^{+0.18}_{-0.28}}$\,Hz/erg, assuming negligible dust attenuation, as expected for $ L<L^*$ LBGs at $z>4$ (e.g., \citealt{dunlop2017,bouwens2021, casey2021}) and from the extrapolation of the results for LAE at lower redshifts  (e.g., \citealt{naidu2022}). If instead we consider a case with a small amount of dust attenuation, we obtain $\xi_{\mathrm{ion},0}=10^{25.84^{+0.18}_{-0.28}}$\,Hz/erg. In deriving this dust-impacted value we assumed, for simplicity, the \citet{calzetti2000} attenuation law and the same $A_V=0.2$\,mag value for both the stellar continuum and nebular emission, given the relative contribution of the two components is still quite uncertain (e.g., \citealt{buat2018, shivaei2020, reddy2020,li2021} and references therein).  These high values of $\xi_\mathrm{ion}$ require young stellar populations (ages $\lesssim10^7$\,yr - e.g., \citealt{robertson2021}), consistent with the values we find from our SED fitting (see also \citealt{stefanon2022a}).

In Figure \ref{fig:csi_z} we compare the value of $\xi_{\mathrm{ion},0}$ from this study to previous estimates at similar redshifts and down to $z\sim2$ (\citealt{stark2015, stark2017, marmol-queralto2016,nakajima2016, bouwens2016, matthee2017, harikane2018, shivaei2018,debarros2019,lam2019,faisst2019,tang2019,nanayakkara2020,emami2020,endsley2021b, naidu2022} and \citealt{atek2022}). We only considered $\xi_\mathrm{ion}$ estimates that refer to samples with either $M_\mathrm{UV}$ within $\pm1$\,mag of the UV luminosity for our stack or $\log M_\star$ within $\pm1$\,dex of the stellar mass we estimated with \textsc{FAST}.

Our measurement is consistent with the estimates existing at $z\sim6.5-8$ (\citealt{stark2015, stark2017,debarros2019,endsley2021b}). Their estimates derived, respectively, from modelling the intense \ion{C}{4}$\lambda 1548$\AA\ identified in the spectrum of a $z=7.045$ galaxy (\citealt{stark2015}), from the [\ion{C}{3}] and Ly$\alpha$ lines in three $z\sim7$ galaxies  with evidence for significant [\ion{O}{3}] emission as suggested by their IRAC colors (\citealt{stark2017}), and from SED fitting to multi-wavelength photometry (\citealt{debarros2019, endsley2021b}). Our measurement is also broadly consistent with the estimates at $z\sim5$ (\citealt{bouwens2016, harikane2018, lam2019,faisst2019} - see also \citealt{maseda2020} for exceptionally high $\xi_\mathrm{ion}\approx10^{26.3}$\,Hz erg$^{-1}$ in lower mass galaxies at $z\sim4-5$). Overall, these results suggest that $\xi_\mathrm{ion}\approx10^{25.6-25.8}$\,Hz/erg could be typical at these epochs, and that  $\xi_\mathrm{ion}\approx10^{25.7}$\ might represent a reasonable estimate.

The  values for $\xi_\mathrm{ion}$ at $z\sim2-3$ are characterized by a large dispersion, ranging $\approx10^{24.7}-10^{26.0}$\,Hz/erg. This distribution could be explained at least in part by selection effects. Remarkably, the values of $\xi_\mathrm{ion}$ from samples characterized by high EW line emission (\textit{low-z analogues} - \citealt{nakajima2016, tang2019}, see also \citealt{chevallard2018} for similar values at $z\sim0$), and from Ly$\alpha$ emitters (\citealt{naidu2022}) are consistent with those found at $z\sim7-8$. Instead, the $\xi_\mathrm{ion}$ estimated from more inclusive samples are generally lower (\citealt{matthee2017, shivaei2018, nanayakkara2020, emami2020, atek2022}). A formal fit to the evolution of $\xi_\mathrm{ion}$, after excluding those from high EW samples, results in $\log(\xi_\mathrm{ion}/\mathrm{[Hz\ erg}^{-1}])=(0.09\pm0.01) z + (24.82\pm0.08)$, whose slope is consistent with predictions from recent models (e.g., \citealt{finkelstein2019} - but see e.g., \citealt{matthee2022} for a non-evolving $\xi_\mathrm{ion}$ model). The high value for $\xi_\mathrm{ion}$ resulting from large EW samples at all redshifts $z>2$ does indicate that we may be settling on a value broadly appropriate for early times for sub-L* LBGs. 

\subsubsection{Implications for $f_\mathrm{esc}$}

\begin{figure*}
\centering\includegraphics[width=12cm]{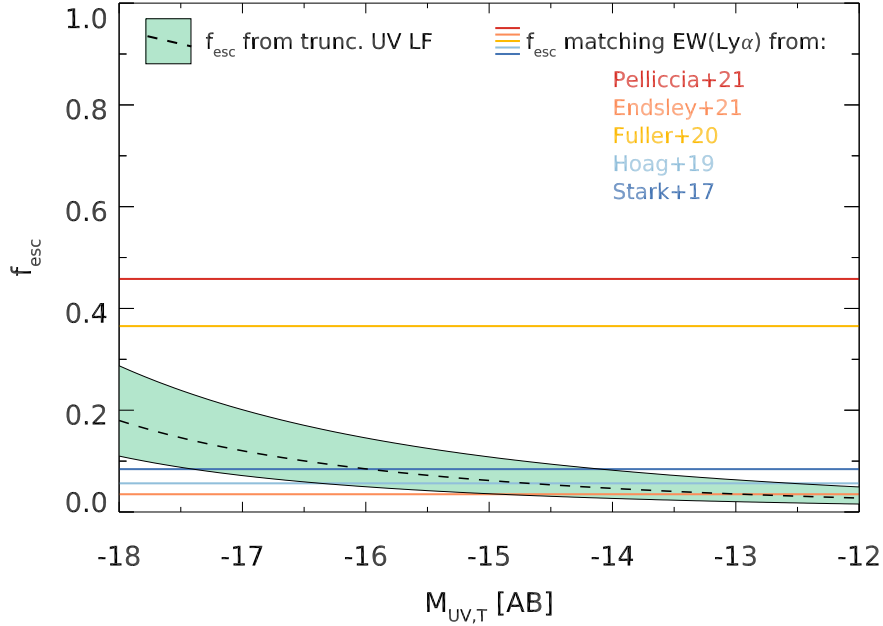}
\caption{The black dashed curve and green shaded area mark the escape fraction and 68\% confidence interval estimated as a function of the UV LF turn-over magnitude $M_T$, as set for the needed ionizing emissivity, and based on the result for $\xi_\mathrm{ion},0$ from this study. The horizontal lines correspond to the $f_\mathrm{esc}$ estimated matching the rest-frame EW$_0$(Ly$\alpha$) inferred from our H$\alpha$ measurement to a compilation of values from the literature. Specifically, we considered the sample averages from \citet{stark2017, hoag2019} and \citet{endsley2021b},  the single $z>7$ galaxy in \citet{fuller2020} sample (C14215A1), and RELICS-DP7  from \citealt{pelliccia2021}.\label{fig:fesc}}
\end{figure*}

The ionizing emissivity is generally expressed as $\dot N(H^0)=\rho_\mathrm{UV}\xi_\mathrm{ion}f_\mathrm{esc}$ (e.g., \citealt{robertson2013}), where $\rho_\mathrm{UV}$ is the luminosity density at rest-frame UV ($\approx1500-1600$\,AA), and $f_\mathrm{esc}$ is the fraction of ionizing photons escaping into the IGM.  Because it is still uncertain whether the faint-end of the $z\sim7-8$ UV LF presents a turn-over (e.g., \citealt{atek2015,livermore2017, bouwens2017, yue2018, bhatawdekar2019, bouwens2021}), here we explore the impact that different values of turnover have on the $f_\mathrm{esc}$. Our estimates are based on the requirement that all the necessary H-ionizing radiation for reionization is generated by stars. For this, we adopted $\dot N(H^0)=10^{50.75}$\,s$^{-1}$Mpc$^{-3}$ (e.g., \citealt{bouwens2015b, finkelstein2019, mason2019, naidu2020}), and $\xi_{\mathrm{ion},0}$ from this study. In the computation of $\rho_\mathrm{UV}$, we approximated the turnover by truncating the $z\sim8$ UV LF of \citet{bouwens2021} at values spanning $-18<M_T<-12$\,mag.

The result of this procedure is presented in Figure \ref{fig:fesc}. The larger value we find for $\xi_\mathrm{ion}$ translates into $f_\mathrm{esc}\lesssim 30\%$ for $M_\mathrm{UV}>-18$\,mag. Values $f_\mathrm{esc}\lesssim20\%$ have been inferred at $z>6$ by recent studies (e.g., \citealt{castellano2017}). In particular, absence of a turn-over in the faint-end slope down to  $M_T\sim -13$\,mag would only require $f_\mathrm{esc}\sim 5-10\%$ to fully ionize the neutral H at $z\sim8$.  These values are consistent with the $f_\mathrm{esc}\sim5-10\%$ inferred for sub-$L^*$  LBGs at $z\lesssim4$ by an increasing number of studies (e.g., \citealt{marchi2017, naidu2018, pahl2021,siana2010, grazian2016, grazian2017, rutkowski2016, steidel2018}). Together they suggest only a marginal evolution of $f_\mathrm{esc}$ with cosmic time for the average galaxy population. Furthermore, and qualitatively, such small escape fraction values can more easily be reconciled with the strong emission lines inferred at rest-frame optical for $z\gtrsim7$ galaxies (EW$_0$([\ion{O}{3}]+H$\beta>1000$\,\AA\ - e.g., \citealt{smit2014, castellano2017, debarros2019, stefanon2019,stefanon2022a, bowler2020, strait2020, strait2021,endsley2021}). The overall consistency is reinforced by considering that photoionization modelling suggests that the production of such strong emission lines already requires very young ($\approx<10^7$\,yr) stellar population ages (e.g., \citealt{inoue2011, wilkins2020}).

An increasing number of studies are identifying Ly$\alpha$ emission in $z\gtrsim7$ galaxies, with rest-frame EW ranging from $\approx 5-20$\,\AA\ to $>100-200$\,\AA\ (e.g., \citealt{pentericci2014, stark2017, hoag2019, fuller2020, endsley2021, pelliccia2021, larson2022}). A direct conversion of our EW$_0$(H$\alpha$) using Case B recombination coefficients ($L_{\mathrm{Ly}\alpha}/L_{\mathrm{H}\alpha}=8.7$) suggests an \textit{intrinsic} EW$_0$(Ly$\alpha$)$_\mathrm{intrinsic}= 517^{+287}_{-244}$\,\AA. Under the assumption that the fraction of escaping LyC photons is approximately similar to that of Ly$\alpha$ emission (e.g., \citealt{steidel2018,izotov2020}), the implied $f_\mathrm{esc}$ ranges between $\lesssim 10\%$ and $\approx40-50\%$ (Figure \ref{fig:fesc}). These estimates are likely upper limits, given the still significant fractions of non-detection particularly for sub-$L^*$ galaxies (e.g., \citealt{pentericci2014, jung2021}).  Thus the preliminary indications from Ly$\alpha$ studies are consistent with the more direct estimates, and reinforce the likely ready availability of adequate reionizing photons from star-forming galaxies for reionization. To give a sense of sensitivity to any dust, a thin $A_V=0.2$\,mag \citet{calzetti2000} dust screen would lower the Ly$\alpha$ flux by $\sim1.7\times$ (EW$_0$(Ly$\alpha$)$= 298^{+165}_{-141}$\,\AA), and increase the requirement on $f_\mathrm{esc}$ by the same factor. Since very low dust absorption is likely, this suggests that any likely levels of dust would not change our conclusions significantly.

\section{Conclusions}
\label{sect:conclusions}

Our analysis of the deepest \textit{Spitzer}/IRAC data available over extragalactic fields for a large sample of $z\sim8$ LBGs has allowed us to detect and measure for the first time the flux in the H$\alpha$ line in the early Universe, and to explore the resulting implications for re-ionization.  Specifically, we obtained this measurement through a median stacking of Hubble and IRAC data for $102$ LBGs initially identified by \citet{bouwens2015} from Hubble imaging over the CANDELS GOODS-N/S, ERS, XDF, CANDELS/UDS and CANDELS/COSMOS fields.  \citet{stefanon2022a} had previously used a similar median stacking procedure to study the main properties of this sample of $z \sim8$ star-forming galaxies as a function of UV luminosity.  These fields have deep coverage in the \textit{HST}/ACS $V_{606}$ and $I_{814}$ and  \textit{HST}/WFC3 $Y_{105}$, $J_{125}$, $JH_{140}$ and $H_{160}$ bands.  Key for our current study are that these fields also have deep \textit{Spitzer}/IRAC mosaics from the GOODS Re-ionization Era Wide Area Treasury from Spitzer (GREATS - PI: Labb\'e; \citealt{stefanon2021c}). These mosaics combine all the relevant observations acquired with IRAC in the $3.6, 4.5, 5.8$ and $8.0\mu$m bands over the GOODS-N/S fields across the full scientific lifetime of \textit{Spitzer}. In particular, the GREATS $5.8\mu$m imaging is the deepest data available at $\approx6\mu$m before \textit{JWST}, and represents a unique opportunity to probe H$\alpha$ at $6.8\lesssim z \lesssim8.7$.

We extracted median flux densities in the IRAC bands after combining the image stamps cleaned from neighbour contamination through \textsc{Mophongo} (\citealt{labbe2006, labbe2010a, labbe2010b, labbe2013, labbe2015}), and recovered total flux densities using the location-specific PSFs from GREATS. Our main results are the following:

\begin{itemize}
\item Our stack results for $102$ galaxies at $z\sim8$ show a $4.3\sigma$ detection in the $5.8\mu$m band, and a red $[3.6]-[5.8]=0.82 \pm 0.27$\,mag color.  
\item Interpreting the excess in the $5.8\mu$m band as due to emission from H$\alpha$, we infer a rest-frame EW$_0$(H$\alpha$)$=1960^{+1089}_{-927}$\,\AA, corresponding to a luminosity of $\log(L_{\mathrm{H}\alpha}/[\mathrm{erg\, s}^{-1}])=42.62^{+0.15}_{-0.23}$. Our result represents the first direct determination of the H$\alpha$ intensity at $z>6.5$. 
\end{itemize}

These results allow us to draw the following conclusions:
\begin{itemize}
\item Comparison of our new EW$_0$(H$\alpha$) measurement with previous determinations at lower redshifts from the literature suggests that the trend of increasing EW$_0$(H$\alpha$) with redshift (e.g., \citealt{faisst2016}) can be extended up to $z\sim8$.
\item After accounting for a $M_\star/L_\mathrm{optical}$ ratio that depends on cosmic time, the observed evolution with redshift of EW$_0$(H$\alpha$) is consistent with the evolution of the specific accretion rate of the dark matter halos, providing further evidence that the star-formation efficiency is at most marginally evolving with cosmic time in the early Universe.
\item Following the formalism of \citet{bouwens2016c}, our new measurement of $L_{\mathrm{H}\alpha}$ implies an efficiency of production of LyC photon $\xi_{\mathrm{ion},0}=10^{25.97^{+0.18}_{-0.28}}$\,Hz/erg. This constitutes one of the largest $\xi_\mathrm{ion}$ estimates at $0<z<8$ for sub-$L^*$ galaxies ($M_\mathrm{UV}\sim-19.8$\,mag, $M_\star\approx10^{8}M_\odot$). While the uncertainties are large, our new measurement is very consistent with previous estimates at similar redshifts, at $z\sim5$, and with those values at lower redshift inferred from samples with significant nebular line emission.  This consistency is not only reassuring but also points to a surprising uniformity across billions of years for star-forming galaxies. 
\item The large value of $\xi_\mathrm{ion}$ we find suggests that escape fractions $f_\mathrm{esc}\lesssim 10\%$ are sufficient for star-forming galaxies to fully ionize the neutral H at $z\sim8$ through escaping LyC radiation.  The small value of $f_\mathrm{esc}$ is consistent with what is seen at lower redshifts $z\sim2-6$ in star-forming galaxies, reinforcing the likelihood that galaxies alone are responsible for reionization. 
\end{itemize}

It is remarkable to step back and realize that this study was enabled by observations in the $5.8\mu$m band, acquired during the first few years of \textit{Spitzer} scientific operations, a decade and a half ago. The present results highlight once again how powerful and pivotal a small telescope like \textit{Spitzer}  has been, especially when able to leverage robust selections made possible with \textit{HST}.  Fortunately, \textit{JWST}/NIRSpec, NIRCam and MIRI combine and enhance the capabilities of \textit{HST} and \textit{Spitzer}, providing the potential for absolutely game-changing science in the coming years.

\begin{acknowledgments} 
MS and RJB acknowledge support from TOP grant TOP1.16.057. PAO acknowledges support from the Swiss National Science Foundation through the SNSF Professorship grant 190079 `Galaxy Build-up at Cosmic Dawn'. The Cosmic Dawn Center (DAWN) is funded by the Danish National Research Foundation under grant No.\ 140.  We also acknowledge the support of NASA grants HSTAR-13252, HST-GO-13872, HST-GO-13792, and NWO grant 600.065.140.11N211 (vrij competitie). GDI acknowledges support for GREATS under RSA No. 1525754. This paper utilizes observations obtained with the NASA/ESA \textit{Hubble Space Telescope}, retrieved from the Mikulski Archive for Space Telescopes (MAST) at the Space Telescope Science Institute (STScI). STScI is operated by the Association of Universities for Research in Astronomy, Inc. under NASA contract NAS 5-26555. This work is based [in part] on observations made with the \textit{Spitzer Space Telescope}, which was operated by the Jet Propulsion Laboratory, California Institute of Technology under a contract with NASA. Support for this work was provideed by NASA through an award issued by JPL/Caltech.
\end{acknowledgments}

\appendix
\section{Validation of the flux density measured in the $5.8\mu$m band}
\label{app:simul}

\begin{figure}
\includegraphics[width=18cm]{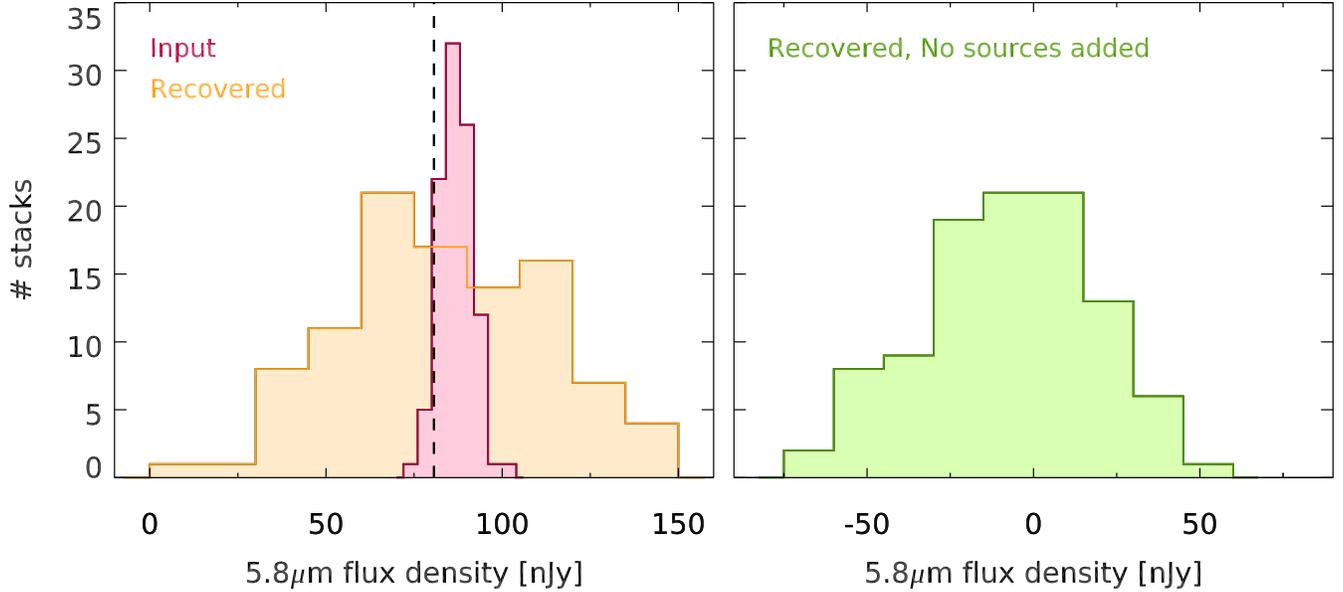}
\caption{{\it Left panel:} Comparison between the distribution of flux densities adopted as input for our Monte Carlo simulations (red histogram) and that after recovering the synthetic sources following the same procedure we adopted for our main analysis (yellow histogram - see also Section \ref{sect:stacking}). The vertical black dashed line indicates the median of our measurements. {\it Right panel:} Distribution of flux densities measured by stacking neighbour-cleaned stamps centered at locations free from existing sources. These two results clearly indicate that our $5.8\mu$m flux density measurement is genuine and any systematics resulting from the imperfect subtraction of neighbouring sources are negligible. \label{fig:simul}}
\end{figure}

We performed a Monte Carlo simulation to ascertain whether the signal detected in the $5.8\mu$m band is genuine emission from the sample of LBGs at $z\sim8$ and not the result of residual contamination from neighbouring sources. We generated a new set of $102$ flux densities in the $H_{160}$ band by randomly scattering the $H_{160}$ measurements of our sample according to their associated uncertainties. We then computed the flux densities in the $5.8\mu$m band by assuming a constant ratio $f_{5.8}/f_{160}=2.2$ between the flux density in the $5.8\mu$m band ($f_{5.8}$) and that in the $H_{160}$ band ($f_{160}$), consistent with what we measure in our stack. This assumption is equivalent to a $z\sim8$ flat $f_\nu$ SED, with a rest-frame EW$_0$(H$\alpha$+[\ion{N}{2}])=2300\AA\ line emission contributing to the $5.8\mu$m flux density. Point sources having the previously computed $5.8\mu$m flux densities were then added at random locations across the four mosaics, adopting the location-specific PSFs from GREATS. In doing this, we preserved the relative fraction and luminosity distribution of sources in each field present in our original sample. Our adoption of point sources is supported by the smaller sizes ($R_e\lesssim1$\,kpc, corresponding to $\approx 0\farcs2$ at $z\sim8$) of sub-$L*$ galaxies at these redshifts (e.g., \citealt{shibuya2015,  bouwens2021b, bouwens2022}), compared to the $5.8\mu$m PSF FWHM ($\approx 2\farcs0$). Following the same procedure we implemented for our main analysis (Section \ref{sect:stacking}), we then constructed neighbour-cleaned stamps using \textsc{Mophongo}, and extracted the photometry of the median stack adopting a $2\farcs0$-diameter aperture, correcting to total using the reconstructed PSF. All these steps were repeated $100$ times. The results of this simulation are presented in the left panel of Figure \ref{fig:simul}. This clearly shows that our analysis is able to recover the median of the input flux densities. To test the amount of systematics that non-optimal removal of neighbouring sources could introduce into our measurements,  we also extracted the photometry from neighbour-cleaned stacks centered on locations free from sources, as inferred from the combination of the $J_{125}$, $JH_{140}$- and $H_{160}$-band mosaics (i.e., this is equivalent to adopting $f_{5.8}/f_{160}=0$). The outcome of this second experiment is shown in the right panel of Figure \ref{fig:simul}. As we might expect if there is negligible contamination from the neighboring sources, the measurements are normally distributed around $0$ nJy. The present Monte-Carlo simulation results significantly increase our confidence in the overall robustness of our $5.8\mu$m-band flux measurements.

\bibliographystyle{apj}


\end{document}